\documentclass{elsart}
\usepackage{graphicx}

\newcommand{\beq}{\begin{equation}}
\newcommand{\eeq}{\end{equation}}

\newcommand{\Cer}{\v{C}erenkov }
\newcommand{\gar}{$\gamma$-ray}
\newcommand{\gars}{$\gamma$-rays}

\begin{document}

\begin{frontmatter} 
\title{The $^{16}$N Calibration Source for the Sudbury Neutrino Observatory}

\author[Orst,LBL]{M. R. Dragowsky\thanksref{LANL}},
\author[Queens]{A. Hamer\thanksref{LANL}},
\author[LBL]{Y. D. Chan},
\author[CRL]{R. Deal},
\author[CRL]{E.D. Earle},
\author[Penn]{W. Frati},
\author[CRL]{E. Gaudette},
\author[Queens]{A. Hallin},
\author[Queens]{C. Hearns},
\author[Lu]{J. Hewett\thanksref{JHewett}},
\author[CRL]{G. Jonkmans\thanksref{Unine}},
\author[LBL]{Y. Kajiyama},
\author[Queens]{A. B. McDonald},
\author[Queens]{B. A. Moffat},
\author[LBL]{E. B. Norman},
\author[CRL]{B. Sur},
\author[UoG]{N. Tagg\thanksref{Oxford}}

\address[Orst]{Oregon State University, Corvallis, OR 97331, USA}
\address[LBL]{Lawrence Berkeley National Laboratory,
Berkeley, CA 94720, USA} 
\address[Queens]{Queen's University, Kingston, Ontario K7L 3N6, Canada}
\address[CRL]{Atomic Energy of Canada Limited, Chalk River Laboratories,
Chalk River, Ontario K0J 1J0, Canada}
\address[Penn]{University of Pennsylvania, Philadelphia, PA 19104-6396, USA}
\address[Lu]{Laurentian University, Sudbury, Ontario P3E 2C6, Canada}
\address[UoG]{University of Guelph, Guelph, Ontario N1G 2W1, Canada}

\thanks[LANL]{Los Alamos National Laboratory, Los Alamos, NM 87545, USA}
\thanks[JHewett]{Department of Physics and Astronomy, University of Western Ontario, London, Ontario  N6A 3K7, Canada}
\thanks[Unine]{Institut de Physique, Universit\'{e} de Neuch\^{a}tel, Neuch\^{a}tel, CH-2000, Switzerland}
\thanks[Oxford]{Nuclear and Astrophysics Laboratory, Oxford University, Keble Road, Oxford, OX1 3RH, UK}
\begin{abstract} 

A calibration source using $\gamma$-rays from $^{16}$N (t$_{1/2}$ = 7.13 s) $\beta$-decay has been developed for the Sudbury
Neutrino Observatory (SNO) for the purpose of energy and other calibrations. The $^{16}$N
is produced via the (n,p) reaction on $^{16}$O in the form of CO$_{2}$ gas
using 14-MeV neutrons from a commercially available  Deuterium-Tritium (DT) generator.
The $^{16}$N is produced in a shielding pit in a utility room near the SNO
cavity and transferred to the
water volumes (D$_{2}$O or H$_{2}$O) in a CO$_{2}$ gas stream via
small diameter capillary tubing.
The bulk of the activity decays in a
decay/trigger chamber designed to block the energetic $\beta$-particles yet permit the primary branch
6.13 MeV $\gamma$-rays to exit. Detection of the coincident $\beta$-particles with plastic
scintillator lining the walls of the decay chamber volume provides a tag
for the SNO electronics.
This paper gives details of the production,
transfer, and triggering systems for this source along with a discussion
of the source $\gamma$-ray output and performance.

\end{abstract}
\end{frontmatter}
\normalsize
PACS:  26.65.+t  Solar Neutrinos; 29.25.Rm  Sources of Radioactive Nuclei; 29.40.Mc  Scintillation Counters; 23.20.Lv Gamma Transitions and Level Energies

Keywords:  Sudbury Neutrino Observatory; Solar Neutrinos; Radioactive Source; Energy Calibration; $^{16}$N $\beta$-decay

Corresponding Author:  M.R. Dragowsky, MS J514, Los Alamos National Laboratory, Los Alamos, NM 87545.  Telephone:  (505) 667-4461, Fax:  (505) 665-4121, Email:  dragowsky@lanl.gov.

\section{Introduction}
\label{intro}
The Sudbury Neutrino Observatory (SNO) is a heavy water \Cer detector designed primarily for the detection of solar neutrinos.  SNO identifies neutrinos via the detection of \Cer light from energetic electrons produced by 
charged current (CC) and elastic scattering (ES) interactions, and neutron capture $\gamma$-rays resulting from the neutral current (NC) 
interactions. It does this with
an array of $\approx$ 9500 photomultiplier tubes surrounding the D$_{2}$O target volume (1 kilo-tonne)
which is contained within a 12 m diameter acrylic vessel. The entire assembly
is immersed within a $\approx$ 7 kilo-tonne water shield and located
under $\approx$ 2100 m of rock in INCO's Creighton mine near Sudbury, Canada.
A complete description of the SNO detector and its many subsystems is presented in
reference~\cite{snonim}. A short description of the SNO energy response will be presented here to provide the context for the discussion of the $^{16}$N calibration source.

A measure of SNO's energy is the number of photomultiplier tubes registering as hit in an event, Nhit.  In addition, the hit pattern (in time and space) provides the information that can be used to reconstruct event interaction position, direction, and energy.  Nhit and the associated hit patterns are dependent not only on the amount of light (number of photons) produced via the \Cer process, but also on the details of the detector optics and the response of the individual photomultiplier tubes. A proper understanding of the light production, propagation, and detection is required in order to understand the response of the detector to neutrinos, muons and radioactive backgrounds. Therefore a detailed Monte Carlo modeling of the detector was developed along with an extensive 
program for optical and energy calibrations. A discussion of the Monte Carlo and the overall program for detector calibration can be found in~\cite{snonim}.

The response of the SNO detector is calibrated using a variety of sources providing isotropic light~\cite{forda,fordb,moffat}, \gars~\cite{hamer,dragowsky,poon,ptg}, $\beta$-particles~\cite{tagg,hamer}, and neutrons. The present work describes one of these sources which provides nearly mono-energetic, primarily 6.13 MeV, \gars\ following the $\beta$-decay of $^{16}$N.  The high-energy \gars\ are used for a number of calibration tasks, including the primary energy scale calibration, verification of the energy resolution and energy scale position dependence, and verification of reconstruction and data reduction algorithms.  A feature essential to the calibration analysis is detecting the $\beta$-particle coincident with the $\gamma$, allowing calibration event identification.  
The $^{16}$N $\beta$-decay is also used as an untagged calibration source for the SuperKamiokande imaging H$_2$O \Cer detector~\cite{skn16nim}.  

The four major systems that comprise the $^{16}$N calibration source are illustrated in Fig.~\ref{fig:msys}. 
The source positioning system is common to all SNO calibration devices and has been discussed in~\cite{snonim}.
The other systems, however, are unique to the $^{16}$N source and are therefore described in this paper.
Earlier discussions of the systems can be found in~\cite{hamer,dragowsky}.
The systems used to produce and transport the short-lived $^{16}$N are discussed in Section~\ref{sec:nprod} and Section~\ref{sec:trans},
respectively.
The design of the decay/trigger chamber (hereafter referred to as the decay chamber), where the $^{16}$N decays and is tagged, is  
described in Section~\ref{sec:design}. The \gar\ output and how it is 
influenced by the decay chamber geometry is discussed in Section~\ref{sec:dcop}.  Finally, in Section~\ref{sec:perf}, the complete source 
performance is demonstrated with laboratory measurements and calibration data obtained in SNO.

\section{$^{16}$N Production}
\label{sec:nprod}
  
SNO uses a commercially available DT generator, the MF Physics model A-320 \cite{mfphysics}, to
produce fast neutrons. A DT generator is a miniature particle accelerator
that generates 14-MeV neutrons by accelerating a mixed beam of deuterium and tritium
onto a target containing both deuterium and tritium. This results in the fusion
reaction
\begin{equation}
d+t\rightarrow n + ^{4}He \;\;\;  (Q \; = \; 17.6 \; MeV)
\end{equation}
where the neutrons produced are emitted nearly isotropically.

The A320 was chosen because of its relatively large neutron flux 
(tunable between 2x10$^{7}$ and 1x10$^{8}$ per second) and the compact
dimensions of the accelerator permitting increased incident flux through the 
production chambers (see below).
The accelerator is cylindrical, 2.2 m long, with a diameter
of 4.29 cm. The DT target is placed 18.73 cm from the end of the accelerator
tube.
It is mounted on a custom built stand and placed
inside a concrete shielding pit located in a lab utility room $\approx$ 40 m away
from the center of the deck above the SNO detector. The pit effectively eliminates any
radiation hazard~\cite{sreport}. 
Fig.~\ref{fig:pit} shows the accelerator and its mounting stand inside the concrete shielding pit.
Also shown are two annular target chambers surrounding the
accelerator DT target used for radioisotope production, and a fast neutron detector for monitoring purposes.
 Both chambers are mounted on a linear table that can be moved remotely to optimize either chamber's position relative to the
DT target. This permits a maximal neutron flux through the chamber being used and hence a maximal
radioisotope yield.

$^{16}$N can be conveniently produced using 14-MeV neutrons via either
the $^{16}$O(n,p) [Q = -9.64 MeV, $\sigma$ = 35 mb], or the $^{19}$F(n,$\alpha$)
[Q = -1.52 MeV, $\sigma$ = 25 mb] reactions~\cite{dunford}.
Liquid targets were initially favored because
the greater mass density would provide much greater $^{16}$N
yields \cite{liquids}. The radioisotope transfer
and the triggering with the $\beta$-particle, however, 
was realized to be much simpler if a gas transfer system was used \cite{sur}.
CO$_{2}$ is therefore used as both target and carrier.
Oxygen-containing (n,p reaction) and 
fluorine-containing (n,$\alpha$ reaction)
fibers were also tested for use in the gas target chambers 
with a variety of carrier gases (He, N$_{2}$, air, etc) \cite{hamer}. 
Certain fibers provided good yields but CO$_{2}$ has proven
adequate.

The target chamber design was a compromise between optimizing the 
$^{16}$N production and  minimizing the residence time of the gas in the chamber (see Section~\ref{sec:trans}), 
as well as constraints imposed by the accelerator and shielding pit configurations. An annular
design was chosen for maximal solid angle coverage. The dimensions of the
annulus were chosen to optimize the number of $^{16}$N atoms produced and entrained in the gas 
flow exiting (or escaping) the chamber.
To estimate the production yield in the chamber a point source for the DT target was assumed.
The yield per isotropically emitted neutron in the chamber volume elements was then
calculated. Solving the integral over these volume elements results in the following
equation for the overall yield~\cite{murn,hamer}
\begin{equation}
\begin{array}{c}
Y_{n}=       \sigma \rho     \left[ \frac{h}{2}
\log \left(1+\left(\frac{R_{2}}{h}\right)^{2}\right)             
-\frac{h}{2} \log \left(1+\left(\frac{R_{1}}{h}\right)^{2}\right)
\right]
-
\\ \\
\sigma \rho \left[ R_{1} \arctan \left(\frac{R_{2}}{h}\right)
-R_{1} \arctan \left(\frac{R_{1}}{h}\right) \right]
+
\\ \\
\sigma \rho (R_{2}-R_{1})
\left[\frac{\pi}{2} - \arctan \left(\frac{R_{2}}{h}\right)  \right]
\end{array}
\end{equation}
where $\sigma$ is the cross section in cm$^{2}$, $\rho$ is the gas density in atoms cm$^{-3}$,
$h$ is the half-height of the chamber, and $R_{1}$ and $R_{2}$ are the
inner and outer radii, respectively. The method for estimating the efficiency for
$^{16}$N to escape the chamber is given in the following section. Optimizing both 
production and escape efficiency, the dimensions were chosen to be a 7.9 cm
length and a 11.44 cm diameter. Fig.~\ref{fig:targ} shows the main features of the final
target chamber design including the target volume and the gas input and output lines. 
Included in the design are distributor plates at the top and bottom of the chamber designed
to promote uniform flow of gas across the chamber.

\section{The Transfer System}
\label{sec:trans}

The purpose of the gas capillary transfer system is to 
transport the short-lived isotopes quickly from the target chambers to the 
deck above the detector, then down to the decay chamber inside
the D$_{2}$O or H$_2$O volume.  For $^{16}$N, CO$_{2}$ from a compressed
gas bottle is sent via polyethylene tubing to a control panel 
that directs the gas at the chosen flow rate and head pressure
 to the target chamber at the bottom of the pit. From there, the 
$^{16}$N produced is entrained in the gas stream and sent
via Teflon tubing to the SNO deck above the detector. 
There, the gas flow is sent down an
umbilical (see Fig.~\ref{fig:umb}) a further 30 m through a thin-walled Teflon capillary to the 
decay chamber placed inside one of SNO's water volumes. The return flow is sent back up to the deck
through a polyethylene line in the umbilical. The capillary and polyethylene are arranged
coaxially. From the deck, the flow is sent back to the control
panel via polyethylene tubing where it can be sent to the lab 
exhaust.

The choice of capillary diameter, target and decay chamber volumes and operating pressure, and flow rate was to
maximize the yield as well as take into consideration the practical constraints imposed by
the source deployment, available capillaries, and gas supply limitations. The overall efficiency is the product of the
target chamber escape efficiency, the transfer efficiency across the capillary,
and decay chamber efficiency.  The methods used to estimate each of these efficiencies as a function of the 
design and operating variables are given below.

To estimate the efficiency for $^{16}$N to escape the target chamber, $\epsilon_{tgt}$, as well as 
the efficiency for $^{16}$N to decay in the decay chamber, $\epsilon_{dec}$,  a full mixing model is
assumed for the gas flow through these volumes (laminar flow models have also been considered). In the full
mixing model, the following balance equation can be used
\begin{equation}
\frac{dN}{dt} = R-\lambda_{N16}N-\frac{N}{V}\frac{dV}{dt}
\end{equation}
where $N$ is the number of radioactive atoms present at any given time, $R$ is the production (or injection) rate
of the radioactive atoms in the target (or decay) chamber volume, $V$.  ${\lambda_{N16}}$ is the mean decay rate for $^{16}$N, and the last term
is the rate at which gas is swept out of the volume.
This general equation can be easily solved for the steady state, $dN/dt = 0$, and using
\begin{equation}
Q=P\frac{dV}{dt} \;\;\; or \;\;\; \frac{1}{V}\frac{dV}{dt}=\frac{Q}{PV}=\lambda_{p}=\frac{1}{\tau_{p}}
\end{equation}
where $Q$ is the mass flow rate, $\lambda_{p}$ and $\tau_{p}$ are the mean volume turnover rate, and mean turnover time, for the gas
in the chambers. The equilibrium number of radioactive atoms inside the volume
then becomes
\begin{equation}
N=\frac{R}{\lambda_{N16}+\lambda_{p}}
\end{equation}
and therefore the injection/production rate is given by
\begin{equation}
R=(\lambda_{N16}+\lambda_{p}){\times}N.
\end{equation}
Using the ratio of the rate at which radioactive atoms exit the target volume, $\lambda_{tgt}N$, and the 
production rate where $\lambda_{p}$ ($\tau_{p}$) is given by $\lambda_{tgt}$ ($\tau_{tgt}$) in this instance 
for Equation 6, the target chamber
efficiency becomes
\begin{equation}
\epsilon_{tgt}=\frac{\lambda_{tgt}}{\lambda_{N16}+\lambda_{tgt}} = \frac{1}{1+\frac{\tau_{tgt}}{\tau_{N16}}}.
\end{equation}
Here $\tau_{tgt}$ is calculated from
\begin{equation}
\tau_{tgt}=\frac{P_{tgt}V_{tgt}}{Q}
\end{equation}
where $P_{tgt}$ is the chamber presssure and $V_{tgt}$ the volume.

The efficiency  for radioactive decay in the decay chamber is taken from
the ratio of the rate at which the radioactive atoms decay inside the volume, $\lambda_{N16}N$, to the injection
rate where in this instance $\lambda_{p}=\lambda_{dec}$ for Equation 6, so
\begin{equation}
\epsilon_{dec}=\frac{\lambda_{N16}}{\lambda_{N16}+\lambda_{dec}}=\frac{1}{1+\frac{\tau_{N16}}{\tau_{dec}}}
\end{equation}
with 
\begin{equation}
\tau_{dec}=\frac{P_{dec}V_{dec}}{Q}.
\end{equation}
In Equation 10, $P_{dec}$ is the decay chamber pressure and $V_{dec}$ is the decay chamber volume.

To calculate the transfer efficiency across the capillary, $\epsilon_{cap}$, 
the transit time is first calculated using
\begin{equation}
\tau_{cap}=\int_{0}^{l} \frac{P(x)}{Q}A dx 
\end{equation}
where $P(x)$ is the pressure along the tube length,
$Q$ is the mass flow rate, and $A$ is the capillary cross sectional area. For $^{16}$N transport, flow
is in the turbulent regime so the average pressure between the target
and decay chambers is used for $P(x)$ as an approximation. 
Given the above, one can estimate the fraction of radioactive gas atoms that survive passing through the
capillary, or the capillary transfer efficiency,  $\epsilon_{cap}$, using
\begin{equation}
\epsilon_{cap}=e^{-\frac{\tau_{cap}} {\tau_{N16}}}.
\end{equation}

Given the results of studies with the above model and the choice of available capillaries and other geometric constraints,
the system dimensions and operating conditions were chosen.
These are summarized in Table ~\ref{tab:tubes} along with other parameters used to estimate the yield.
Also given are the results of sample 
calculations for the various efficiencies and yield.
Note that the primary transfer line consists of two segments, $l_{1}$ and $l_{2}$, with cross sectional areas $A_{1}$ and $A_{2}$, 
respectively.
The second segment is used inside the umbilical and $P_{mid}$ is the pressure at the junction of the two segments.      
The sample calculation using these values predicts that the yield
should be roughly 460 $^{16}$N  triggers per second. It should be noted that this estimate is only approximate because
of the various assumptions made, for example complete mixing in the target and
decay chambers. Also, the ranging out of the $^{16}$N recoils in the target 
chamber is not taken into account, and finally slight deviations of the
capillary diameters (few $\%$) can strongly affect the calculated yield. Still, the calculations suggest that sufficient
amounts of $^{16}$N should be produced for the chosen design parameters.

\section{$^{16}$N  Decay Chamber Design}
\label{sec:design}

The $^{16}$N  decay chamber design results from a compromise between the need to minimize $\gamma$-ray attenuation and maximize $\beta$-particle containment.  The design is illustrated in Fig.~\ref{fig:n16design}. The main casing is a smooth cylindrical tube of stainless steel that is 41.9 cm long, 10.16 cm in diameter and has a wall thickness of 0.476 cm.  The stopping power in steel is 2 MeV cm$^{2}$/g and 1.6 MeV cm$^{2}$/g for 10 MeV and 3-4 MeV $\beta$-particles, respectively.  Therefore, the wall thickness selected for use in the decay chamber, together with the scintillator and sleeve material (described below), will greatly reduce the flux of emerging $\beta$-particles.  

The main casing achieves a gas seal using stainless steel top and bottom plates having grooves along the circumference to hold O-rings that seat on the inner surface of the tube.  An interior annular plate with O-ring separates the chamber into upper and lower volumes.   A 0.159-cm thick sleeve of stainless steel (aluminum) maintains a constant distance between the interior annular plate and the bottom (top) plate.  The tube interior surface is threaded for a length of 1 cm at each end to accommodate threaded annular plates that back up the top and bottom plates, so that the O-rings cannot move when subjected to the decay chamber operating pressure.  The casing diameter is constrained to fit within guide tubes that penetrate the deck in SNO, allowing calibration source deployment into the region between the PMT Support Structure (denoted PSUP in Fig.~\ref{fig:msys}) and the acrylic vessel.  The sealing design incorporating O-rings allows the maximum diameter casing for deployment through the guide tubes.

The $^{16}$N  decays occur within a region bounded by a 3 mm thick cylindrical shell of plastic scintillator (Bicron BC400~\cite{bicron1},~\cite{bicron2}) located in the lower volume.  The scintillator is 15.24 cm in height and has outer diameter 8.89 cm.  Gas is delivered to the decay chamber via the umbilical described in Section~\ref{sec:trans}.  The polyethylene return line is secured to a fitting atop the inner annulus.  The inner Teflon capillary extends through the interior of the scintillator, causing $^{16}$N  gas to be introduced at the bottom of the plastic scintillator.  The outlet for the gas is the polyethylene return line tubing.  A uniform distribution of gas and a long dwell time in the scintillator are promoted by introducing gas to the bottom and removing it from the top of the scintillator.  

The upper volume contains a 5.08 cm diameter photomultiplier tube (Electron Tubes, Ltd. model 9208B~\cite{etl}) that monitors the scintillator for light through an optical coupling in the interior annular plate.  The PMT high voltage is generated using a high voltage converter (Spellman model MHV12-2.0K1000N~\cite{spellman}) that is collocated with the PMT.  The high voltage converter requires 12 V input (obtained from the umbilical) and produces a maximum output of 2,000 V.  Normal operations call for 1,200-1,500 V.  The top plate has an aperture in the center to allow coupling to the umbilical.  The umbilical-to-top plate connection is sealed externally using a sequence of three O-rings and pressure plates.  Space is provided above the PMT in the upper volume for coupling of these items to their counterparts. 

The optical coupling between the upper and lower volumes is achieved with a rigid acrylic window that occupies the central portion of the interior annular plate.  An aluminum ring that makes an O-ring seal with the interior annular plate is used to mount the acrylic window.  Coupling of the window to the scintillator is via a 0.318 cm thick optical pad.  The window is coupled to the PMT by another optical pad.  The PMT is secured using two springs that exert a force toward the optical pad.  This arrangement allows for the PMT or scintillator to be independently separated from the optical coupling window.  It also maintains a consistent alignment of the relative positions of the PMT and scintillator. 

\section{$^{16}$N Decay Chamber Function and $\gamma$-Ray Emission}
\label{sec:dcop}

A simplified $^{16}$N decay scheme is presented in Fig.~\ref{fig:n16ds} \cite{toi}.  The branch of primary interest for calibration produces a $\beta$-particle with end point energy 4.3 MeV and a 6.1-MeV \gar\ (66.2\%).  There are other branches that produce \gars\ in coincidence with $\beta$-particles  (6\%).  There is also a direct branch to the ground state, resulting in a 10.4-MeV endpoint $\beta$-particle without an associated \gar\ (28\%).  Thus, each $\beta$-particle will produce a trigger for SNO, but not all triggered events constitute a valid calibration event.  

The weak $\beta$-decay branch to the 7.1 MeV excited state introduces an uncertainty into the mean $\gamma$-ray energy triggered by the source.  This is due to the readout threshold imposed on the energy deposited in the thin-walled scintillator.  Two distinct energy deposition modes (dE/dx and energy spectroscopy) compete to shift the mean energy either higher or lower, respectively.  In the dE/dx case, energy deposition is greater for lower energy $\beta$-particles, increasing the representation of the 7.1-MeV $\gamma$-ray in the mean.  Alternately, energy deposition proportional to $\beta$-particle energy favors the 6.1-MeV $\gamma$-rays, since that $\beta$-spectrum is harder.  Neither effect is expected to be large.  The combined effect is evaluated through stepping the readout threshold through the scintillator spectrum during source characterisation in the SNO detector.

The $^{16}$N \gars\ must transit the decay chamber material before emerging into the heavy water of SNO.  The following is a list of issues concerning the effect by the decay chamber material on the \gar\ emission spectrum.
\begin{itemize}
\item \gar\ attenuation from interactions in the material.
\item \gar\ production following $\beta$-particle bremsstrahlung in the material.
\item isotropy of the \gar\ emission spectrum.
\end{itemize}

The \gar\ emission spectrum is reduced in number and downgraded in energy relative to the production spectrum by the steel.  The Monte Carlo emission spectrum is presented in Fig.~\ref{fig:emission}.  The $^{16}$N decay scheme \gars\ are denoted in the figure using vertical arrows.  The \gar\ emission spectrum features a continuum background that must be accounted for when analyzing data from this source.

Energetic $\beta$-particles may interact in the decay chamber material causing secondary \gar\ emission from the source due to brems\-strah\-lung.  These \gars\ will generate \Cer\ photons that either sum with those from the primary 6.1-MeV \gar, or generate independent events associated with the decay to the ground state.   The effect of brems\-strah\-lung on the $^{16}$N energy response has been studied in Monte Carlo and is found to be small.  This result is expected because the cross sections for multi-MeV photon emission are themselves small.  Inner bremsstrahlung in the decay of $^{16}$N has been neglected, as it is expected to be below $10^{-3}$ in probability per decay.

The decay chamber geometry is cylindrically symmetric.  It is therefore expected that the \gar\ emission spectrum will feature anisotropy in the polar angle.  The SNO detector also exhibits a polar angle anisotropy due to the penetration through the PMT Support Structure (PSUP) by the Acrylic Vessel (AV) chimney.  The SNO detector anisotropy in polar angle is more restricted than that of the $^{16}$N decay chamber, therefore, the anisotropy in the \gar\ emission spectrum will be due primarily to the decay chamber material.

Clearly, the chamber geometry will affect the energy output of the source.  However, particle transport simulation concerning electromagnetic interactions in materials is included in the SNO Monte Carlo (EGS4~\cite{egs4}) and should properly take into account these effects.   Also, our analysis is not completely dependent on Monte Carlo methods.  Neutron calibration sources have been developed by SNO that provide 6.25-MeV \gar\ data from neutron capture on deuterium unaffected by source materials.  This data has been used as a cross reference in our calibration program to assess the decay chamber influence on the \gar\ emission spectrum.

Detailed studies have been done by the SNO collaboration to assess the additional uncertainty the above source features introduce into the energy calibration.  A full discussion is beyond the scope of this paper; therefore, the conservative overall uncertainty estimate is presented here as being less than 0.5\%.

\section{Source Performance}
\label{sec:perf}

The following discusses the performance of the $^{16}$N source based on both off-line measurements and measurements taken in SNO. The discussion focuses on the yield and the trigger efficiency.

The yield is primarily a measure of how well the production and transfer systems work.  The method for estimating the yield was outlined in Section ~\ref{sec:nprod} and ~\ref{sec:trans}  and the predicted yield was $\approx$ 460 per second. To test these calculations, measurements were taken with the SNO transfer configuration and the optimal flow conditions. The rate was taken as the chamber PMT trigger rate above the measured background for the given PMT threshold. The maximum measured rate was 300 decays per second. This is 35 $\%$ less than the predicted yield but this is not surprising given the assumptions outlined in sections 3 and 4.  These rates have proven adequate for the intended SNO calibration purposes.  A single-point energy calibration with statistical precision better than 0.1\% can be obtained in about ten minutes.  Reconfiguration of the transfer lines (constrained at present by the source deployment hardware) would permit greater yields but this is not necessary at present.

Given the geometry of the decay volume and the surrounding scintillator, the energy of the tagging $\beta$-particles, as well as the care taken to optimize the light collection of the system, a high trigger efficiency was expected. This was confirmed via off-line measurements in which the signal from the $^{16}$N  source chamber was used to trigger a nearby 12.7 cm diameter by 12.7 cm length NaI detector.  Singles, coincidence and background spectra were obtained and analyzed to determine the trigger efficiency in the range 6 -- 7 MeV range.  The result was 95$\pm$2\%.~\cite{dragowsky}.  Results in SNO are consistent with this and this is illustrated in Fig.~\ref{fig:zf} where the number of position reconstructed events that trigger the chamber PMT are shown along with those that do not. As shown, almost all the events that reconstruct near the source chamber are triggered. In this instance the chamber was placed near the bottom of the SNO D$_{2}$O volume and the rate of untriggered events from decays in the umbilical tube is almost constant along the central axis of the detector.  Fig. 9 shows reconstructed X and Z coordinates for untriggered and triggered events with the source located at the center of the D$_2$O.

To illustrate the intended use of the source in SNO, the triggered Nhit distribution for the source at the center of the detector is shown in Fig.~\ref{fig:nmcdata}. Also shown is the Monte Carlo predicted Nhit distribution. In this instance, a single constant representing the overall detection efficiency of the photomultiplier tubes in the Monte Carlo was tuned until the best agreement was achieved in the Nhit means.  Also, for both data and Monte Carlo, a requirement was made that the events reconstruct (i.e. in position) so that ground state $^{16}$N decays in which no light was produced in the D$_{2}$O (i.e. false triggers) could be eliminated. Fig. ~\ref{fig:nmcdata} shows that in this instance, data and Monte Carlo agree quite well.

\section{Conclusion}

$\indent$ A calibration system using the decay of $^{16}$N has been developed and 
successfully deployed into the SNO detector. The source has achieved all its 
design goals including high $^{16}$N production yield ($\ge$ 300 s$^{-1}$) and  high trigger
efficiency ($\ge$ 95$\%$).  The $\gamma$-ray emission from the source is also well understood via Monte Carlo simulation.  The systematic uncertainty in the energy calibration due to using this source is less than 0.5\%. 

\begin{ack}
The SNO project has been financially supported in Canada by the Natural Sciences
and Engineering Research Council, Industry Canada, National Research Council of
Canada, Northern Ontario Heritage Fund Corporation and the Province of Ontario, in
the United States by the Department of Energy, and in the United Kingdom by the
Science and Engineering Research Council and the Particle Physics and Astronomy 
Research Council. The heavy water has been loaned by AECL with the cooperation
of Ontario Power Generation. The provision by INCO of an underground site is greatly
appreciated.

Finally, the authors would like to thank the SNO Collaboration and support staff.
\end{ack}

\clearpage

\begin{figure}
\includegraphics[height=10.cm]{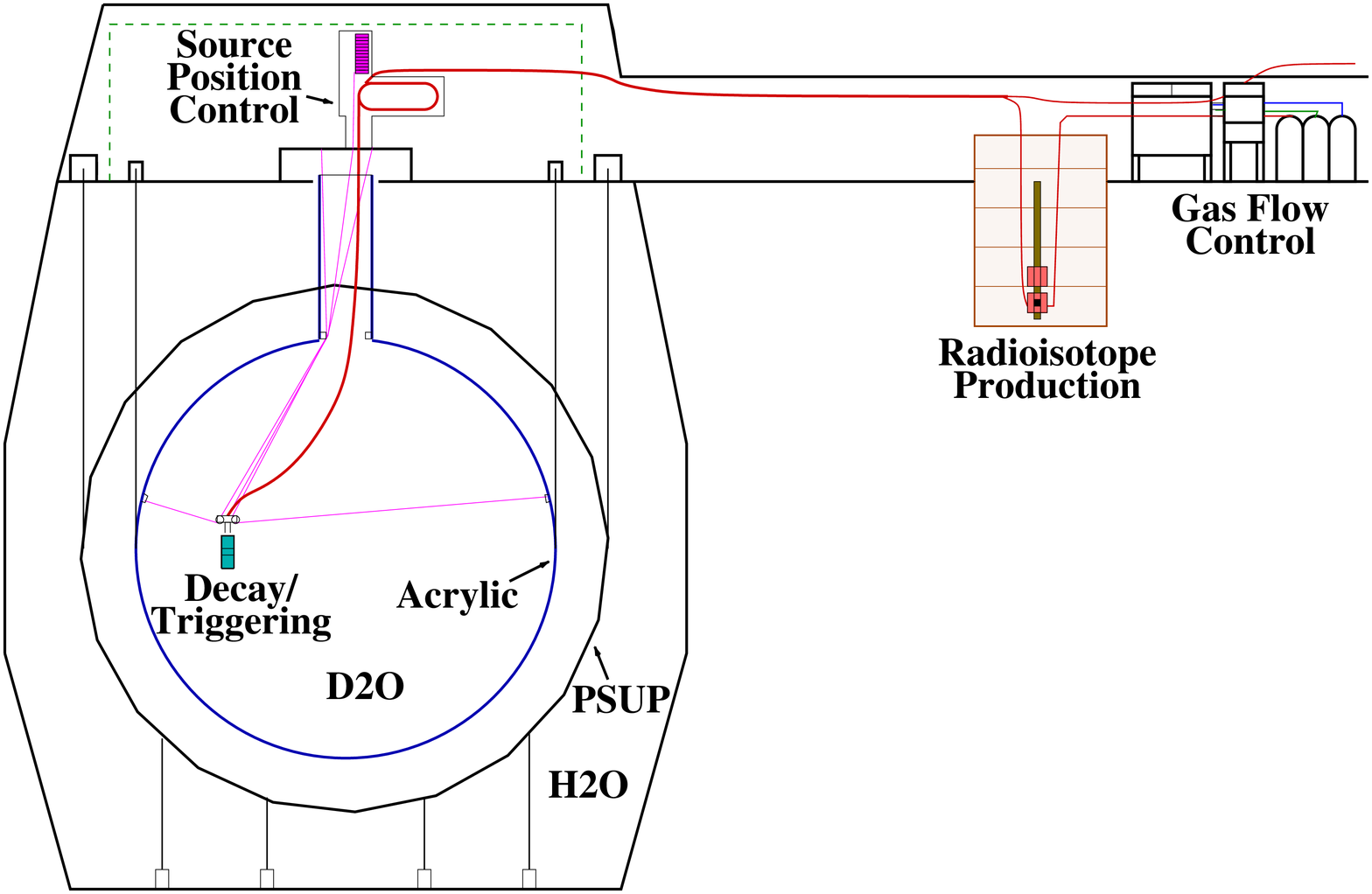}
\caption{}
\label{fig:msys}
\end{figure}

Fig.~\ref{fig:msys}  The basic features of $^{16}$N calibration. Gas Flow Control directs the target and transfer gas.  Radioisotope Production is where the $^{16}$N is produced.  Source Position Control positions the sources within the D$_{2}$O or H$_{2}$O volume.  Decay/Triggering is where $\gamma$-rays exiting into the water volumes are tagged by the detection of the coincident $\beta$-particles.  The Photomultiplier Support Structure is denoted PSUP.

\begin{figure}
\includegraphics*[height=9.cm]{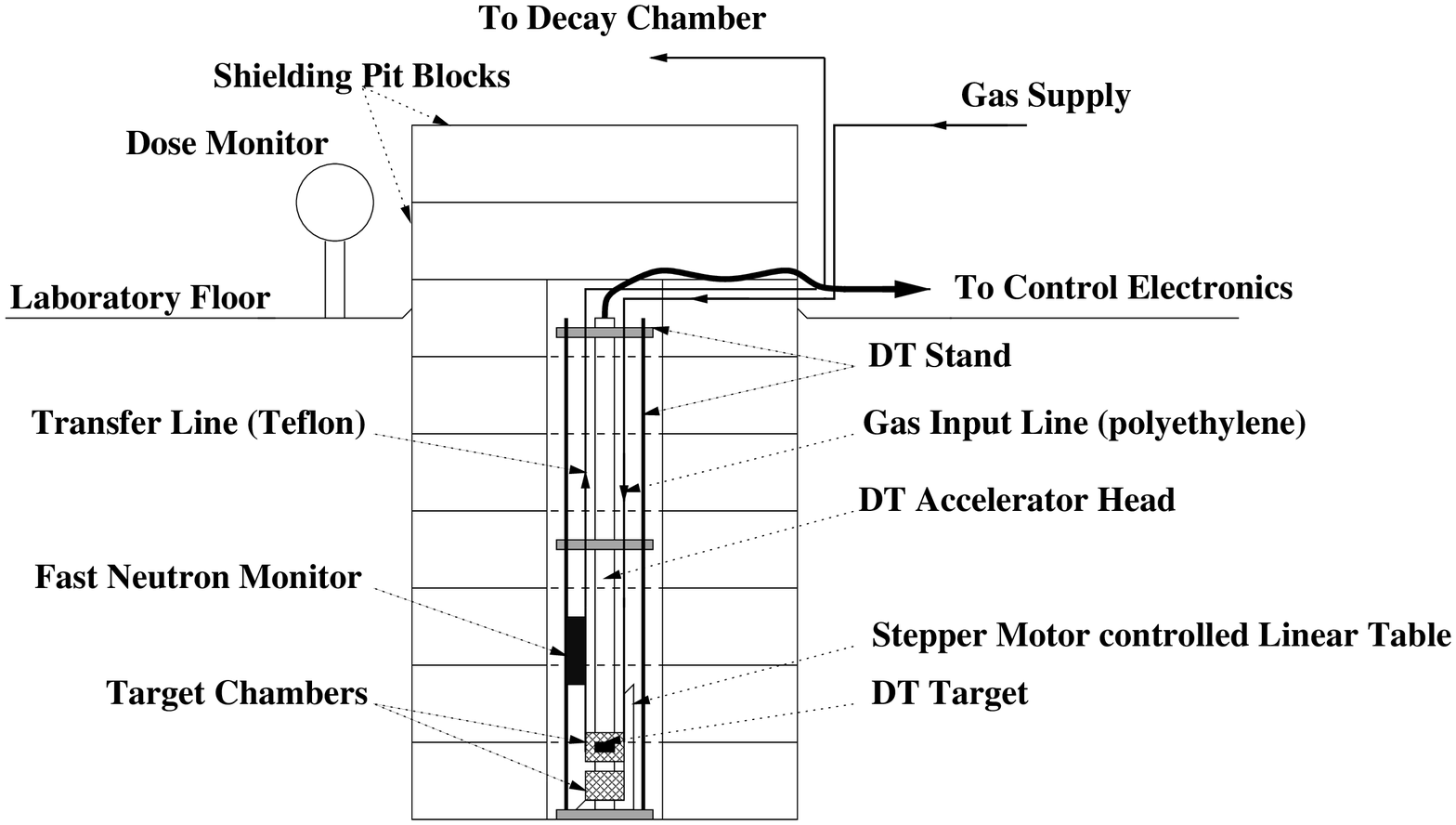}
\caption{}
\label{fig:pit}
\end{figure}

Fig.~\ref{fig:pit}  Contents of the DT generator pit.

\begin{figure}
\includegraphics[height=15.cm]{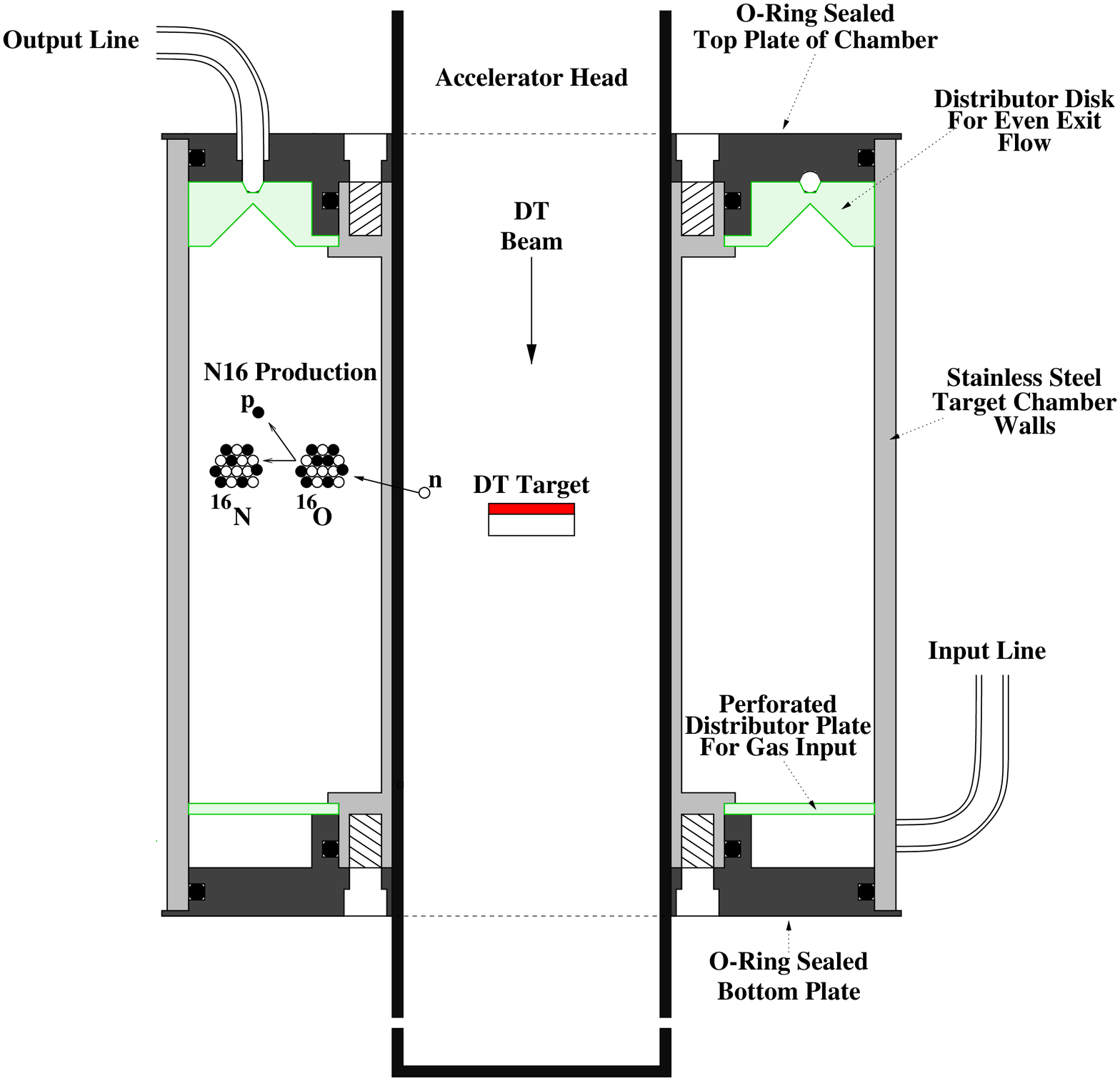}
\caption{}
\label{fig:targ}
\end{figure}

Fig.~\ref{fig:targ}  Cross sectional view of the annular $^{16}$N target chamber.

\begin{figure}
\includegraphics[width=15.cm]{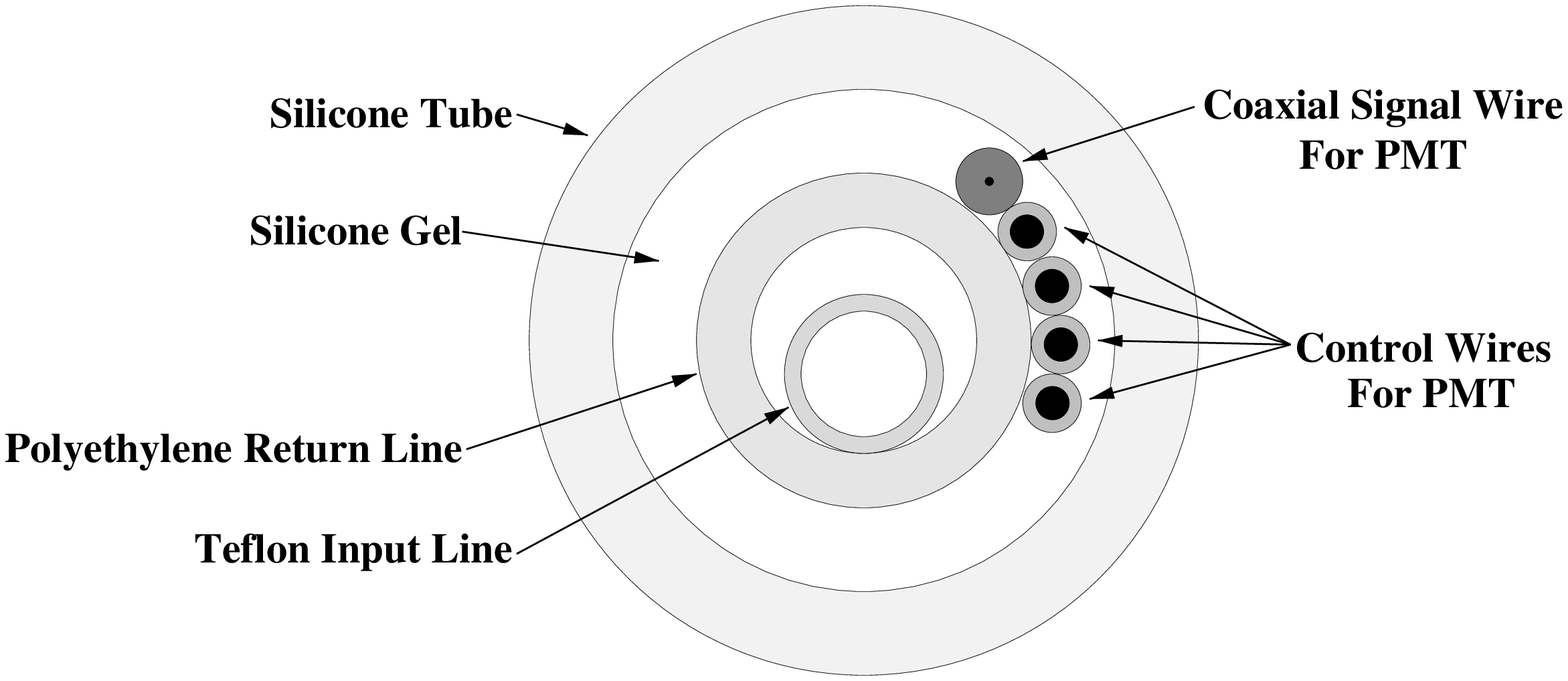}
\caption{}
\label{fig:umb} 
\end{figure}

Fig.~\ref{fig:umb}  Cross section of the $^{16}$N umbilical.

\begin{figure}
\begin{center}
\includegraphics[height=18.cm]{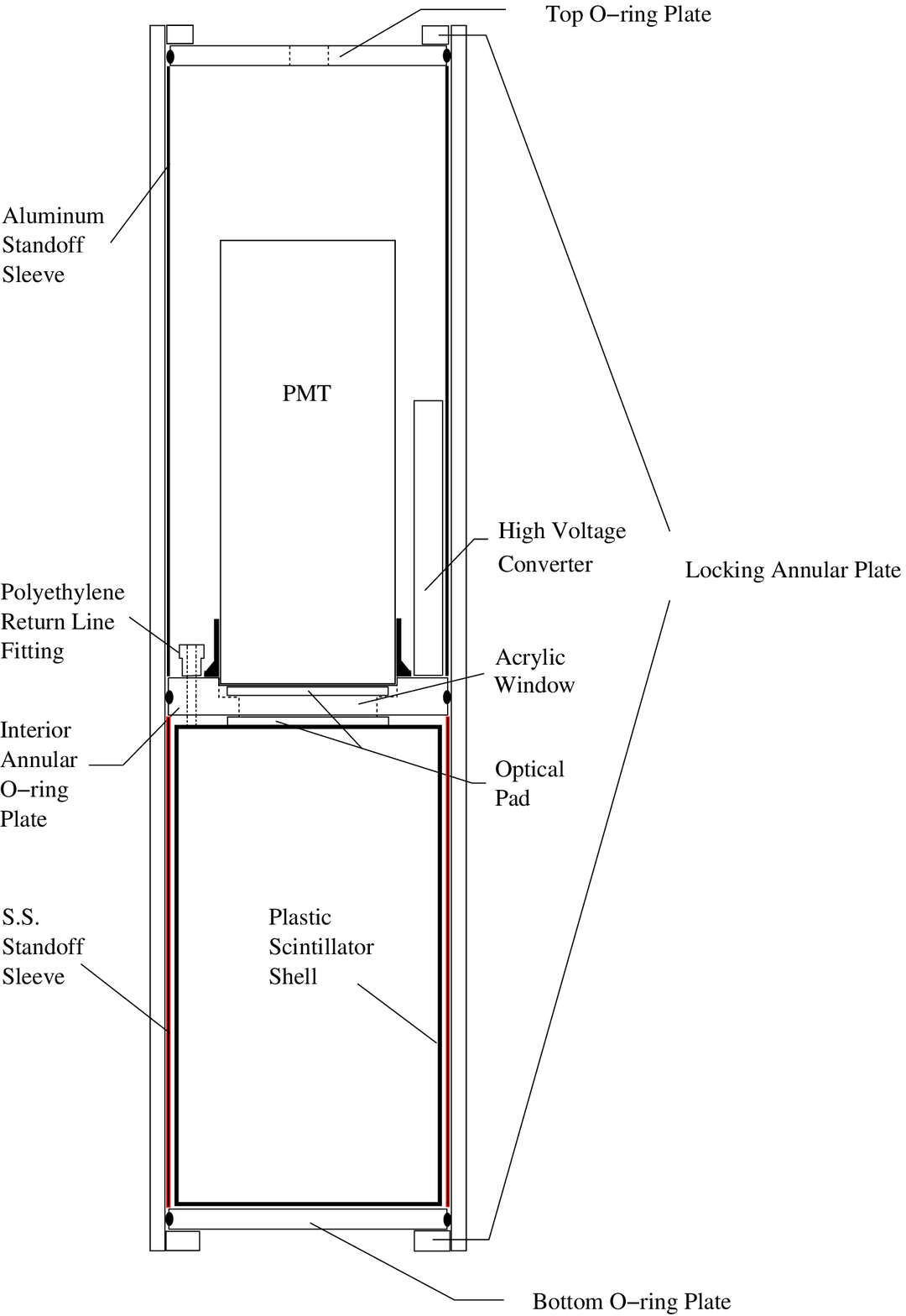}
\caption{}
\label{fig:n16design}
\end{center}
\end{figure}

Fig.~\ref{fig:n16design}  A schematic representation of the decay chamber.

\begin{figure}
\includegraphics[height=10.cm]{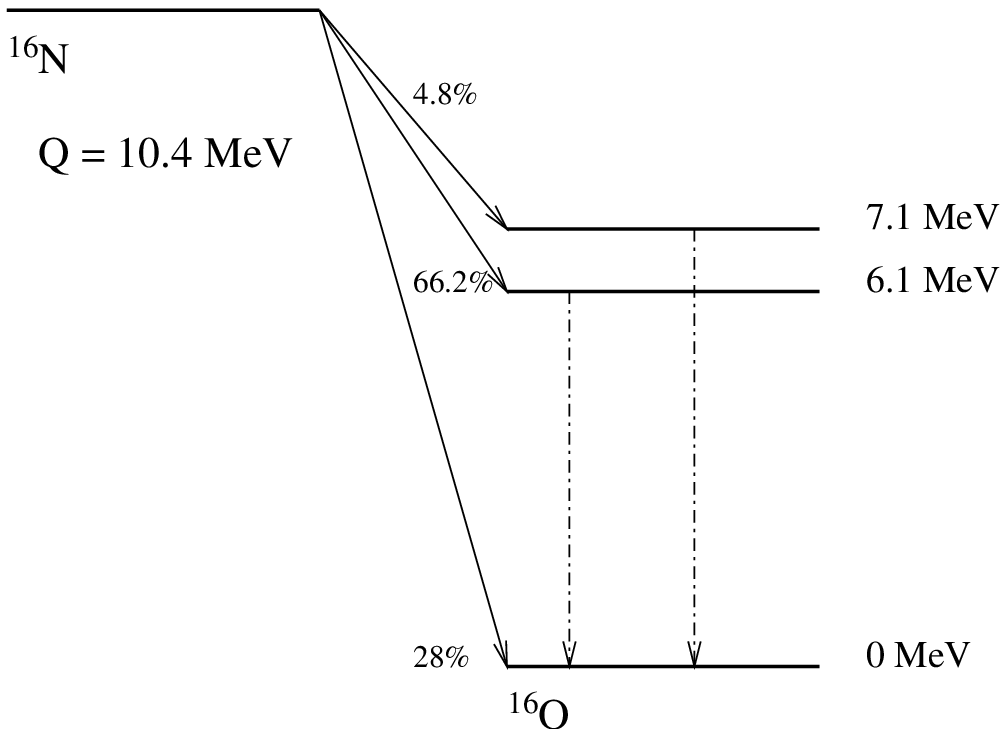}
\caption{}
\label{fig:n16ds}
\end{figure}

Fig.~\ref{fig:n16ds}  Major $^{16}$N  $\beta$-decay branches.  The \gar\ energies appear to the right of the energy levels.  The $\beta$-decay branch strengths are shown as percentages near the arrow depicting the transition.

\begin{figure}
\begin{center}
\includegraphics[height=15.cm]{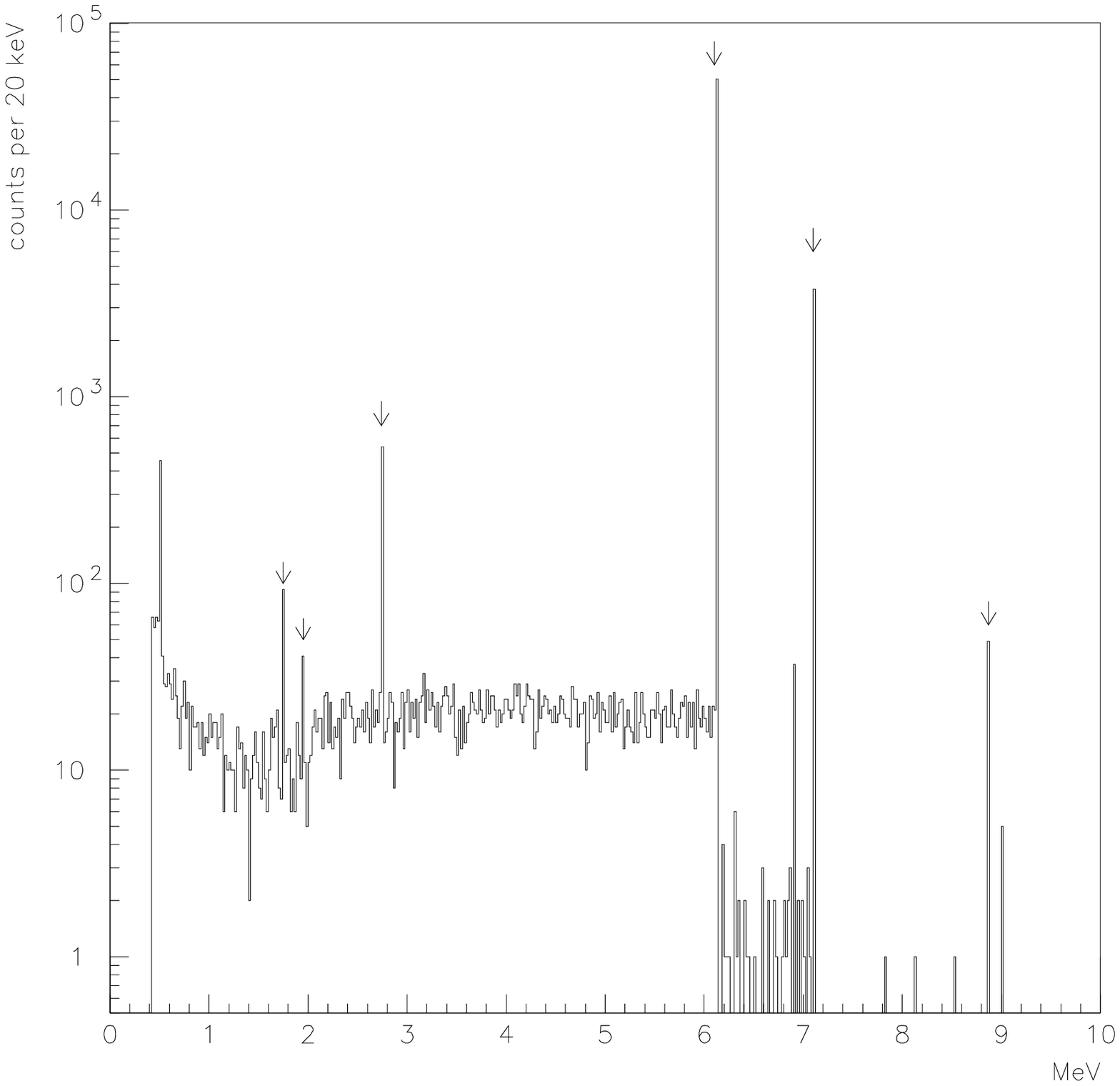}
\caption{}
\label{fig:emission}
\end{center}
\end{figure}

Fig.~\ref{fig:emission}  Monte Carlo simulation results showing the source chamber effect on the \gar\ emission spectrum.  The $^{16}$N $\beta$-decay \gars\ are denoted in the spectrum by downward pointing arrows.

\begin{figure}
\includegraphics[height=10.cm]{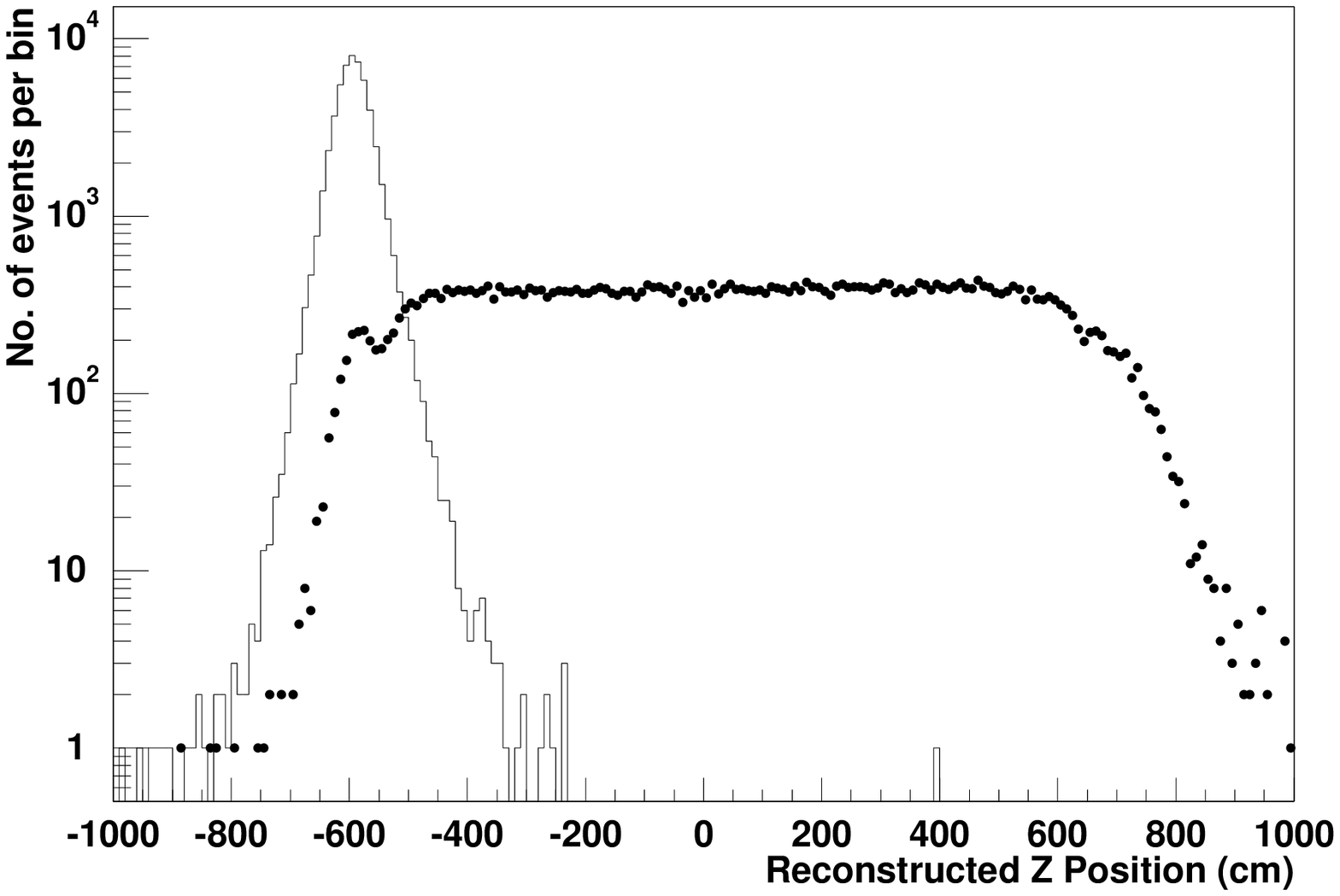}
\caption{}
\label{fig:zf}
\end{figure}

Fig.~\ref{fig:zf}  Reconstructed Z position for triggered events (line) and untriggered events (dots).

\begin{figure}
\includegraphics[height=8.cm]{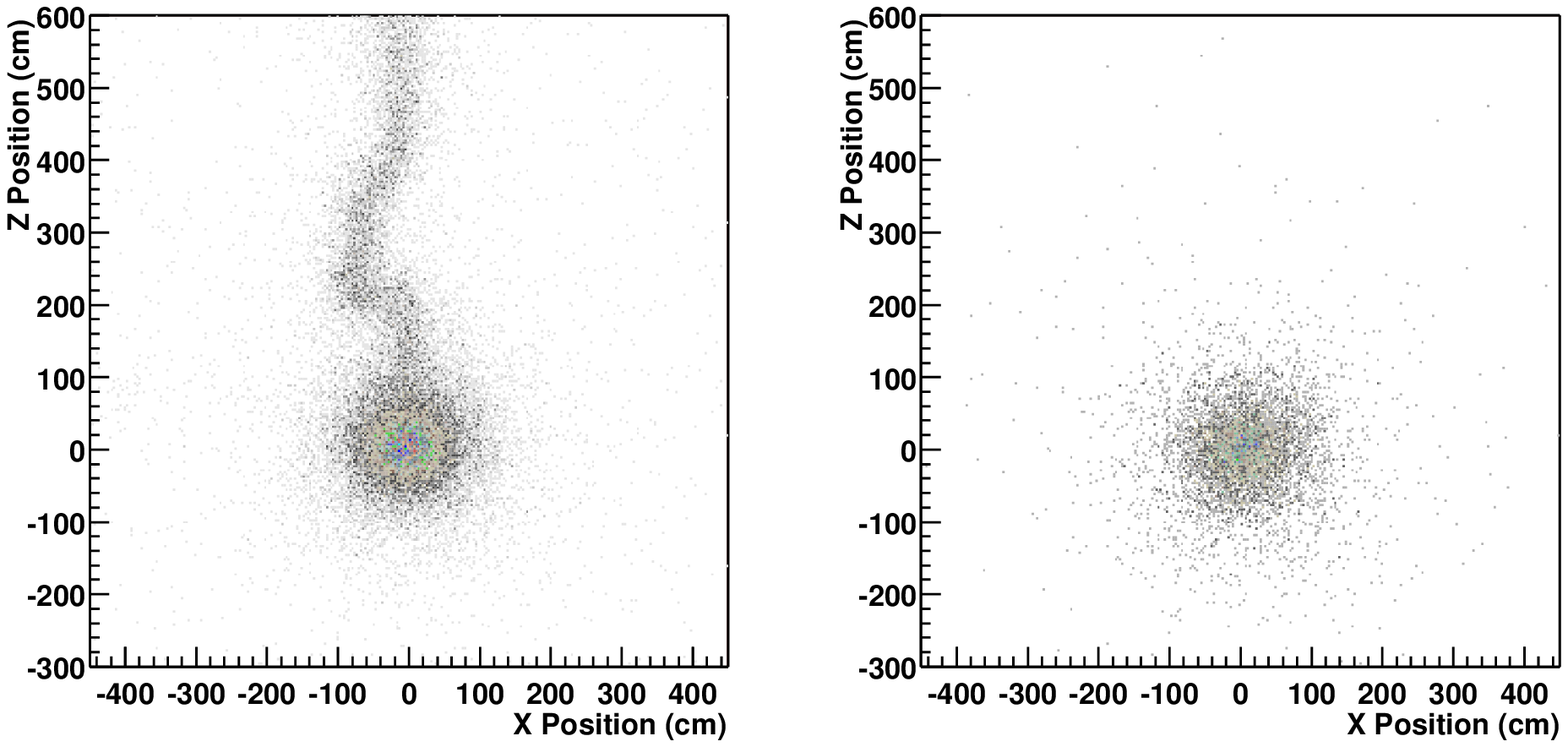}
\caption{}
\label{fig:xy}
\end{figure}

Fig.~\ref{fig:xy}  Reconstructed X and Y positions for the $^{16}$N chamber placed at the center of the
SNO detector. The left panel shows all events including those due to umbilical decays and the
right panel those that are in coincidence with the chamber beta detection.
A cable is used to support the source and not the umbilical, hence the bend in the number of reconstructed
untagged events relative to the vertical axis.

\begin{figure}
\includegraphics[height=10.cm]{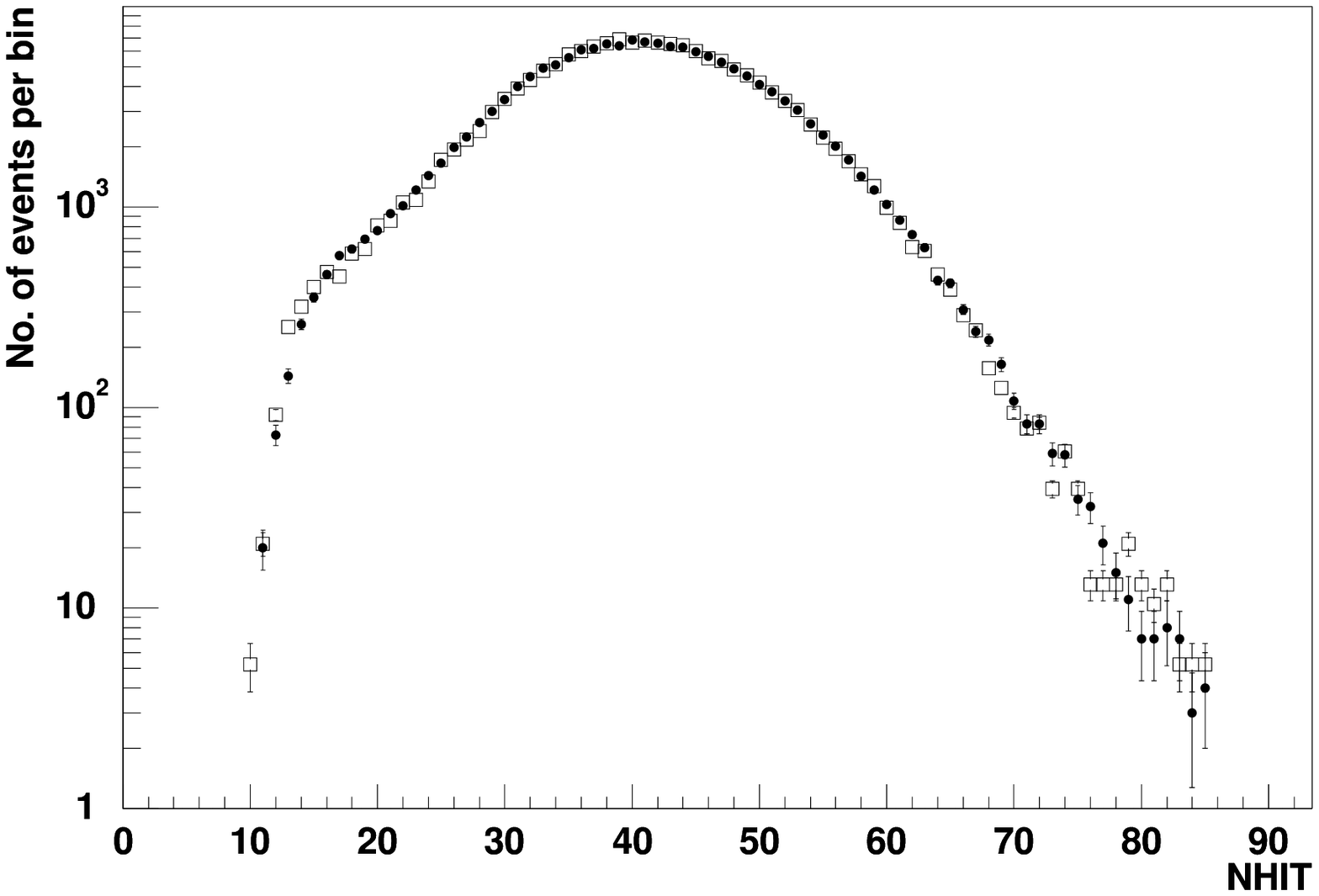}
\caption{}
\label{fig:nmcdata}
\end{figure}

Fig.~\ref{fig:nmcdata}  Comparison of NHIT distributions for the $^{16}$N scaled Monte Carlo (open squares) and data (dots) 
for the source at the center position. The Monte Carlo PMT collection efficiency input was tuned  so as
achieve the 
best agreement with the data for the  Nhit mean.

\clearpage

\clearpage

\clearpage

\clearpage

\clearpage

\clearpage

\clearpage

\clearpage

\clearpage

\clearpage

\clearpage


\begin{table}
\centering
\begin{tabular}{|l|c|} 
\hline
neutron flux                                     &      10$^{8}$ s$^{-1}$                      \\
cross section ($\sigma$)	                 &	35 mb \cite{dunford}                   \\
\hline
Target Chamber pressure P$_{tgt}$                &      6.5 Atm                                \\
Target Chamber half height ($h$)		 &	3.95 cm                                \\
Target Chamber inner radius ($R_{1}$)	         &	2.32 cm                                \\
Target Chamber outer radius ($R_{2}$)	         &	5.72 cm                                \\
Target Chamber volume ($V_{tgt}$)                &      678 cm$^{3}$                           \\
Target Chamber gas density at 6.5 Atm  ($\rho$)  &	3.52 x 10$^{20}$ cm$^{-3}$             \\ 
\hline
Main transfer line length ($l_{1}$)              &      43 m                                   \\
Main transfer line area  ($A_{1}$)               &      0.0793 cm$^{2}$                        \\
Intermediate transfer line pressure ($P_{mid}$)   &     6.1 Atm                                \\
Umbilical transfer line length ($l_{2}$)         &      30 m                                   \\
Umbilical transfer line area ($A_{2}$)           &      0.0455 cm$^{2}$                        \\
\hline
Decay Chamber volume ($V_{dec}$)                 &      1050 cm$^{3}$                          \\
Decay Chamber pressure ($P_{dec}$)               &      4.05 Atm                               \\
\hline
Mass Flow Rate ($Q$)                             &      230 Atm-cm$^{3}$ s$^{-1}$              \\
\hline
$^{16}$N production yield ($Y_{n}$)              &      6.92x10$^{-5}$ $^{16}$N n$^{-1}$       \\
Transfer efficiency ($\epsilon_{tgt}$)           &      34.8 $\%$                              \\   
Transfer efficiency ($\epsilon_{cap}$)           &      29.8 $\%$                              \\
Decay Chamber efficiency ($\epsilon_{dec}$)      &      64.2 $\%$                              \\
Total efficiency ($\epsilon_{tot}$)                                 &      6.7 $\%$                               \\
\hline
Calculated Yield                                 &      460 s$^{-1}$                           \\
\hline
\end{tabular}
\vspace{0.5cm}
\caption{\label{tab:tubes}{System dimensions and parameters for yield estimation. Also shown are the calculated efficiencies and yield.}}
\end{table}

\end{document}